# Spicules in IRIS Mg II Observations: Automated Identification

Vicki L. Herde[1,2], Phillip C. Chamberlin[2], Don Schmit[3], Souvik Bose[4,5,6,7], Adrian Daw[8], Ryan O. Milligan[9], and Vanessa Polito[4,5,10]

[1] University of Colorado Boulder, Boulder, CO, 80303, USA
[2] Laboratory for Atmospheric and Space Physics, 3665 Discovery Drive, Boulder, CO, 80303, USA
[3] Cooperative Institute for Research in Environmental Sciences, 216 UCB Boulder, CO, 80309, USA; Phil.Chamberlin@lasp.colorado.edu
[4] Lockheed Martin Solar & Astrophysics Laboratory, Palo Alto, CA, 94304, USA
[5] Bay Area Environmental Research Institute, NASA Research Park, Moffett Field, CA, 94035, USA
[6] Institute of Theoretical Astrophysics, University of Oslo, PO Box 1029, Blindern NO-0315, Oslo, Norway
[7] Rosseland Centre for Solar Physics, University of Oslo, PO Box 1029, Blindern NO-0315, Oslo, Norway
[8] Solar Physics Laboratory, NASA Goddard Spaceflight Center, 8800 Greenbelt Road, Greenbelt, MD, 20771, USA
[9] Queens University Belfast, University Road, Belfast, BT7 1NN, UK
[10] Department of Physics, Oregon State University, 301 Weniger Hall, Corvallis, OR, 97331, USA
*Received 2022 December 8; revised 2023 February 17; accepted 2023 February 21; published 2023 April 5*

## Abstract

We have developed an algorithm to identify solar spicules in the first ever systematic survey of on-disk spicules exclusively using Mg II spectral observations. Using this algorithm we identify 2021 events in three Interface Region Imaging Spectrograph (IRIS) data sets with unique solar feature targets spanning a total of 300 minutes: (1) active region, (2) decayed active region/active network, and (3) coronal hole. We present event statistics and relate occurrence rates to the underlying photospheric magnetic field strength. This method identifies spicule event densities and occurrence rates similar to previous studies performed using Hα and Ca II observations of active regions. Additionally, this study identifies spicule-like events at very low rates at magnetic field intensities below 20 G, and increasing significantly between 100 and 200 G in active regions and above 20 G in coronal holes, which can be used to inform future observation campaigns. This information can be be used to help characterize spicules over their full lifetimes, and compliments existing Hα spectral capabilities and upcoming Lyα spectral observations with the Solar eruptioN Integral Field Spectrograph (SNIFS) sounding rocket. In total, this study presents a method for detecting solar spicules exclusively using Mg II spectra, and provides statistics for spicule occurrences in the Mg II h line with respect to the magnetic field strength for the purpose of predicting spicule occurrences.

*Unified Astronomy Thesaurus concepts:* Solar spicules (1525); Solar chromosphere (1479); Solar transition region (1532); Solar atmosphere (1477); Solar coronal heating (1989); Solar magnetic fields (1503); Solar active regions (1974); Solar coronal holes (1484); Quiet sun (1322)

## 1. Introduction

One of the major challenges in solar physics today is explaining why the Sun's corona is so much hotter than its visible surface and lower atmospheric layers. This is also known as the "coronal heating problem". There are many sources that may contribute to the energy content of the corona, with magnetic reconnection and wave interaction being the major contributors, though the exact proportions of their contributions are unknown (Klimchuk 2006; De Moortel & Browning 2015).

The Interface Region Imaging Spectrograph (IRIS; De Pontieu et al. 2014) was launched in 2013 and has been providing detailed spectral and spatial information with high spatial and spectral resolution observations of the chromosphere and the transition region (TR). Using data from IRIS, scientists have been able to study a wide range of events such as large-scaled flares and active regions, and small-scaled spicules and network jets. With these observations, we can gain a better understanding of small-event energy and mass contributions to the solar corona as well as the energetics and dynamics within the chromosphere and TR.

The present study focuses on one such small-scaled feature called a spicule. Spicules are small, ubiquitous, jet-like features in the solar chromosphere, which, when observed on the limb, look like slender threads sticking out from the Sun's surface (Roberts 1945). Physically, they are upflowing or downflowing streams of plasma constrained within small magnetic flux tubes above the Sun's surface. They were first identified on the solar limb by Secchi (1871, 1877), and have since been matched to their disk counterparts called rapid blueshift/redshift excursions (RBEs/RREs; Langangen et al. 2008; Rouppe van der Voort et al. 2009; Sekse et al. 2012). RREs and RBEs are so named because they appear to redshift or blueshift the spectral line wing for a short period of time. Ever since their discovery in 1871, spicules have formed a part of numerous reviews such as by Beckers (1968, 1972), Sterling (2000), and more recently Tsiropoula et al. (2012).

Observationally, spicules are divided into two types. Type I spicules, also called dynamic fibrils or mottles, are longer in duration (3–5 minutes). Dynamic fibrils appear primarily in active regions (Pereira et al. 2012), while mottles are their quiet-Sun counterparts (Rouppe van der Voort et al. 2007). They are formed due to the leakage of photospheric oscillations (*p*-modes), which steepen into shocks in the chromosphere (De Pontieu et al. 2004; Hansteen et al. 2006). Type II spicules on the other hand are shorter-lived (1–3 minutes), highly dynamic, and appear more often in regions of quiet Sun and coronal holes (De Pontieu et al. 2007; Pereira et al. 2012). Unlike Type Is, the origin of Type II spicules is still debated, with magnetic reconnection (Ding et al. 2011), the nonlinear propagation of Alfvénic waves (Matsumoto & Shibata 2010), and the release of amplified magnetic tension through ambipolar diffusion (Martinez-Sykora et al. 2017) being the prime candidates.

Type II spicules are often found to be heated to TR (e.g., Pereira et al. 2014; Rouppe van der Voort et al. 2015) and even coronal temperatures (e.g., De Pontieu et al. 2011; Henriques et al. 2016; Samanta et al. 2019). This property makes them a potential candidate that can energize the outer atmosphere of the Sun. To understand how spicules contribute to the energy and mass flow into the corona, we need to understand both how they transfer mass and energy from the photosphere to the corona, as well as their statistical distribution including where they happen, how often they occur, what their durations are, and their spatial extent. This study looks at one currently missing and critical piece of this energy and mass transfer: small-scale, frequently occurring transient events in the chromosphere where Mg II is formed. This is above the Ca II and H$\alpha$ layers where spicules have often been observed, and lower than the TR Si IV 1402 Å which is imaged by IRIS. Thus far, there are no large-scale statistical studies of this type.

As spicules evolve, their changing temperature shifts them into and out of different elemental spectral emission lines. H$\alpha$ and Ca II emissions form lower in the chromosphere and at cooler temperatures while Mg II h 2803.52 Å and k 2796.34 Å emissions and hydrogen Ly$\alpha$ (hereafter called Ly$\alpha$) 1215.67 Å form higher in the chromosphere and at higher temperatures (Leenaarts et al. 2013). This means if a spicule is heated, it is first detectable at H$\alpha$ and Ca II wavelengths, and may then switch into emitting at Mg II, Ly$\alpha$, or even TR Si IV wavelengths if heated enough. To understand a spicule's evolution fully, we need to observe it over its entire lifetime and at multiple heights and temperatures. Such understanding can only be gained through observations across multiple wavelengths.

Statistical information on spicule distribution has been calculated in multiple studies (Pereira et al. 2012; Sekse et al. 2012; Bose et al. 2021, 2023). These studies were primarily performed using H$\alpha$ observations around 6562.81 Å and Ca II K observations around 8542 Å, while Bose et al. (2023) used H$\beta$ around 4861 Å. These authors used data from the Crisp Imaging Spectropolarimeter (CRISP; Scharmer et al. 2008) and the CHROMospheric Imaging Spectrometer (CHROMIS; Scharmer 2017) instruments on the Swedish Solar Telescope (SST; Scharmer et al. 2003), and the Solar Optical Telescope (SOT; Tsuneta et al. 2008) on board the Hinode spacecraft. These observations primarily view emission and absorption effects from plasma in the lower chromosphere and do not extend to hotter plasmas in the TR or corona; therefore, to study the full connection between the chromosphere and corona, we need to study the aforementioned upper chromospheric and TR emissions and absorptions. To this end, we propose a method for identifying spicules in IRIS Mg II h spectral observations and provide statistical information for the events identified, as well as relate these to the photospheric magnetic field intensities derived from the Helioseismic and Magnetic Imager (HMI; Scherrer et al. 2012) on board the Solar Dynamics Observatory spacecraft (SDO; Pesnell et al. 2011). To complement these observations, the upcoming Solar eruptioN Integral Field Spectrograph (SNIFS) sounding rocket mission will be able to observe both spatially and spectrally in a passband surrounding the Ly$\alpha$, O V, and Si III emission lines (Chamberlin & Gong 2016). While there have been relatively few spectral Ly$\alpha$ observation of on-disk spicules so far (Chintzoglou et al. 2018), the Mg II line formation vertically overlaps with Ly$\alpha$ and can provide similar location information (Schmit et al. 2017). Determining the magnetic field strengths that have an abundant occurrence of spicules will enable us to identify potential on-disk targets for the five minute rocket flight of SNIFS, which has a field-of-view (FOV) of 0.5 square arcminutes. This will increase the chances of observing the highest number of such events with Ly$\alpha$.

This paper will first discuss the observations and data used, then cover how spicules affect the Mg II line and how we can algorithmically detect this. We group identifications into spatial and temporal events and analyse their occurrence with relation to the line-of-sight (LoS) magnetic field. All events identified using Mg II spicule spectral parameters are referred to as spicules. We then discuss how Type I and Type II spicules present differently, and provide statistics with respect to location, time, and magnetic field. Finally we discuss the limitations of this method, suggest solutions to these limitations, present future work, and summarize our findings.

## 2. Methods

### 2.1. Observations and Data

The IRIS instrument is a space-based UV slit spectrograph and imager which was launched in 2013. It records spectra in multiple bandpasses including around the Mg II h&k lines as well as cospatial images at multiple wavelengths such as the 1400 Å Si IV bandpass. It has multiple observation modes, including sit-and-stare, which observes only one spatial dimension, and rastering across an image, which collects data in two spatial dimensions (see De Pontieu et al. 2014 for more details). Both modes also record the spectral dimension.

In order to understand the occurrence of spicules in a variety of environments on the Sun, we identified three IRIS data sets spanning the different types of magnetic environments found on the Sun: active region near the disk center, decayed network in the southern hemisphere, and equatorial coronal hole. The details of the data sets used are listed in Table 1. All data sets are located near the disk center to minimize the effects of viewing angle, maximize spicule LoS Doppler effects, and utilize more reliable magnetic field information. We use both the sit-and-stare and high-frequency raster modes. The observations were chosen in a way such that the temporal cadence of the rasters is less than 25 s, which would allow detections of short-lived spicular events. Increased temporal resolution comes at the expense of spatial coverage, but a temporal analysis provides us with a better method for verifying our results than a spatial analysis. Results from a single raster frame of a high-frequency raster are directly comparable to a sit-and-stare frame. The coronal hole data set was co-observed with the Hinode spacecraft and SST, and the decayed network data set was co-observed with Hinode SOT. These external data sets may be useful for further analysis, but that analysis is beyond the scope of this study.

We chose Mg II h over the k line because of the presence of the Mn I and Fe I lines near the k line's wing at 2795.64 and 2795.83 Å, respectively. Since spicules in the form of RBEs and RREs were first identified on-disk using the Mg II line wing, the presence of these two lines could potentially interfere with this analysis.

**Table 1**
IRIS Mg II Data Sets Used

| Data set | Size | Date | Time (UT) | Duration (min) | Obs. ID | Obs. Center (Solar-$X$ and -$Y$) | IRIS Rotation (degrees) | Raster FOV | Raster Cadence | Raster Steps |
|---|---|---|---|---|---|---|---|---|---|---|
| Active Region | large | 11 Nov 2015 | 13:59:14 | 60 | 3600104017 | $-185''$, $-235''$ | 0 | $4'' \times 119''$ | 12.7 s | 4 |
| Decayed Network | large | 14 Feb 2021 | 6:40:32 | 59 | 3660259103 | $48''$, $-376''$ | 0 | $0\rlap{.}''35 \times 119''$ | 9.4 s | 1 |
| Coronal Hole | very large | 24 Sep 2014 | 7:49:34 | 170 | 3820257466 | $78''$, $-167''$ | 90 | $4'' \times 174''$ | 21.6 s | 4 |

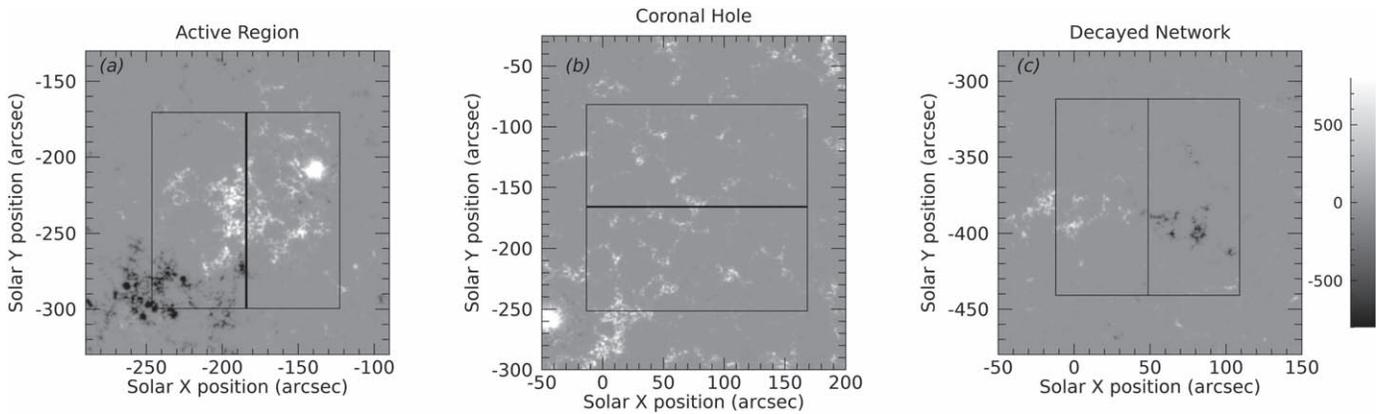

**Figure 1.** The surrounding magnetic field as measured by HMI for each data set, for context. The box in each image shows the IRIS slit jaw imager FOV, while the line in the center shows the location of the slit. Magnetic strength values are in G.

In addition to the Mg II data, LoS magnetic field observations from the HMI were used. For each data set, a single 720 s HMI LoS magnetogram closest to the start time of the IRIS observation was obtained. Separately, the HMI data sets were compared between the beginning and the end of the IRIS observation periods to verify that the magnetic field does not change over the observed duration. 82% of the locations displayed less than a 20 G difference. Schmit & De Pontieu (2016) found HMI LOS data to be poorly correlated with IRIS radiances below 20 G, while the Stanford HMI website lists HMI's precision to be 10 G.[11] Locations with a difference of greater than 20 G will be discussed later in this paper. 720 s was chosen for signal-to-noise ratio purposes. The data set was then interpolated to match the pixel sizes and locations of its corresponding IRIS data set. Figure 1 displays the location of the IRIS slit for each data set with the surrounding magnetic field strength for context.

### 2.2. Spicule Features in Mg II

Rouppe van der Voort et al. (2015) and Bose et al. (2019, 2021) have identified spectral signatures of spicules in Mg II, Hα, and Ca II. Bose et al. (2021) identified spicules in the Hα absorption line by selecting profiles which displayed a center-of-gravity shift greater than 20 km s$^{-1}$, a peak difference between intensity profiles of at least 0.2, and a line core width of at least 1.2 Å. Rouppe van der Voort et al. (2015) and Bose et al. (2019) list spicule signatures in Mg II to include a Doppler shift of the k3 feature, a dimming of the k2 peak in the direction of the k3 Doppler shift, and a brightening in the wing between the k1 and k2 features in the Doppler-shifted direction without strong dimming in the other wing. Typical ranges for some of these signatures are provided, but no minimum requirements. Figure 2(a) shows the names ascribed to features of an Mg II h spectrum and Figure 2(b) provides a visual example of spicule signatures. We build on this work by translating these signatures into quantifiable measurements of the Mg II line.

---

[11] http://hmi.stanford.edu/Description/hmi-overview/hmi-overview.html

### 2.3. Selection Parameters

To capture the unique spectral features of on-disk spicules systematically, we propose several parameters which can be used together to identify spicular activity in the Mg II line. In order to do this, we first make a simplifying assumption that the Mg II line can be modeled as the sum of a positive emission and a negative absorption Gaussian similar to Schmit et al. (2015). Figure 2(c) shows an example of a positive and negative Gaussian which combine to fit an Mg II spectrum.

The simplest way to identify the Doppler shift of the absorption feature would be to find the minimum value between the two h2 peaks, but in many of the most Doppler-shifted spicule spectra, one of the h2 peaks is entirely suppressed, meaning there is no minimum value. Instead, we use the Doppler location of the absorption Gaussian. We represent the change in h2 peak intensities by using the ratio between the h2r and h2v peaks. Using a ratio allows us to identify changes in the line regardless of the overall line intensity. Finally, we use a combination of measurements of the emission Gaussian to represent the asymmetric wing enhancement. We capture the asymmetry of the feature using the Doppler shift of the emission Gaussian requiring that it be in the same direction as the shift in the absorption Gaussian. We also require that the Gaussian's width ($\sigma$) be broader than normal so that the nonbrightened wing remains relatively unaffected. Figure 2(c) visually highlights the Doppler location of the emission and absorption Gaussians, as well as the width ($\sigma$) of the emission Gaussian.

### 2.4. Identifying Spicule Spectra

The IRIS data were first smoothed vertically along the slit using three pixel boxcar smoothing to increase the signal-to-noise ratio. Then the spectra were modeled using the Gaussian fitting method described above to determine the absorption Gaussian centroid Doppler shift, the emission Gaussian centroid Doppler shift, and the emission Gaussian width. While these fitted parameters do not necessarily correlate to physical properties of the plasma due to the optically thick nature of the line, they can be used for identifying generic changes in the line shape.

This work will be used by the SNIFS rocket when making observation targeting decisions, which require a single estimate over the entire Sun near the center of the disk regardless of the magnetic regime. With this in mind, we estimate baseline values against which we will compare each spectrum by calculating the mean of the means between data sets, and the mean of the standard deviations between data sets for each parameter listed in Section 2.3. We chose a mean-of-means because the coronal hole data set has an order of magnitude more data points than the others, meaning it would dominate a simple mean.

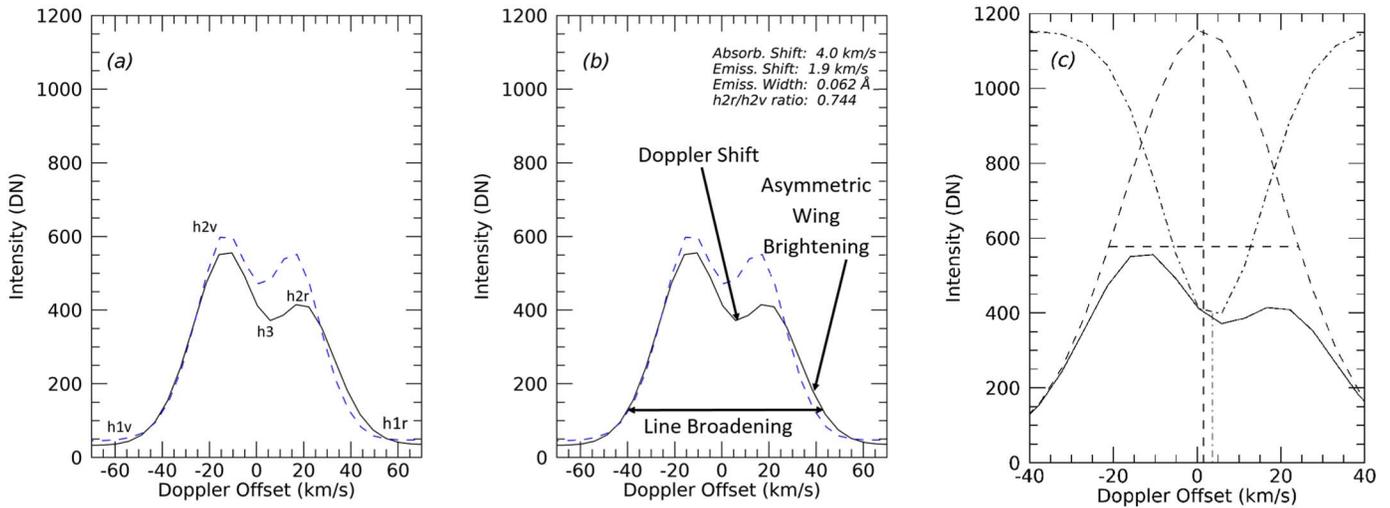

**Figure 2.** Example of a redshifted spicule under the IRIS slit. In subplot (a) the h1–h3 labels call out features in the Mg II line. The solid black line is the measured spectrum while the blue dashed line is a reference spectrum, found by taking the median of all three data sets. Subplot (b) highlights the ways a spicule influences the Mg II spectrum. Parameters used to identify this spicule are listed at the top right. Subplot (c) displays the positive Gaussian (dashed) and negative Gaussian (dotted–dashed) combined to fit an Mg II spectrum, as well as visually displaying the width and central peak locations derived from their associated Gaussians.

We designate a series of thresholds, listed in Table 2 and visually represented in Figure 3, which a spectrum must satisfy to qualify a measurement as an event. To calculate the threshold values, the means and standard deviations for each parameter within each data set were calculated. Then the average values between the three data set means (mean-of-means) and standard deviations (mean-of-standard-deviations) were found. To identify positive events, the individual fitted values for each measured spectrum were compared to their respective mean-of-means. The peak ratio standard deviations for h2r/h2v and its inverse were calculated separately since ratio standard deviations are asymmetric. Spectra lacking one of their h2 peaks were marked as passing the ratio parameter, while spectra entirely lacking an absorption feature were marked as failing the absorption Doppler shift parameter.

To be marked as a positive event, a spectrum needed each parameter to be greater than 0.68 standard deviations from the mean, while also having its absorption and emission Gaussians share a Doppler shift direction. Because of the optically thick nature of the Mg II line, the emission Gaussian width and Doppler shift do not directly correspond to physical quantities. For this reason, we do not select a specific physical cutoff value to indicate an event, and instead used the mean and standard deviation of the data sets. This means that different data sets will have slightly different cutoff values when calculated independently. The value of 0.68 was chosen because it selected half of the spectra for each parameter as a potential event candidate. This does not guarantee that half of our spectra will be identified as events, as an event must satisfy this requirement for all four parameters. A co-observational comparison to previously identified Hα or Ca II spicules will be required to refine this number further.

**Table 2**
Threshold Values for Positive Event Identification

| Identification Parameter | Threshold for Positive Events. |
|---|---|
| Absorption Doppler Shift | less than $-0.623$ km s$^{-1}$ or greater than $+1.514$ km s$^{-1}$ |
| Emission Broadening | greater than 0.0467 Å |
| Emission Doppler Shift | less than $-0.765$ km s$^{-1}$ or greater than $+0.912$ km s$^{-1}$ |
| Peak Ratio (Red/Blue) | less than 0.916 or greater than 1.314 |

**Note.** The Doppler shift values are not symmetric due to the overall redshifts of the data sets. The lower ratio value was calculated as blue/red and converted to red/blue due to asymmetries in calculating the standard deviations of the ratios.

In the case of these three data sets, the threshold values for each parameter are listed in Table 2. Figure 3 shows the distribution of values for the selection parameters in each data set. Subplots (a), (b), and (d) show that the peak ratio, emission Doppler shift, and absorption Doppler shift values follow a roughly Gaussian distribution. Subplot (c), however, shows that the active region and decayed network emission broadening distributions are not Gaussian, but are instead bimodal. This skew is strongly influenced by the underlying magnetic field strength, and will be addressed in Section 4.

### 2.5. Identifying Discrete Events

In order to understand the spatial and temporal information about the identified events, we aggregated regions of identified spicule pixels neighboring in space and time into single events. Many event regions appeared to be connected but were separated by one pixel containing a nonspicule spectrum. To connect these regions, two-pixel boxcar smoothing was performed over the entire data set both in time and space including corner neighbors. The event data set was a trinary data set where redshifted events had a value of 1, blueshifted events had a value of $-1$, and nonevents had a value of 0, meaning the results after the smoothing had a range from $-1$ to 1. Regions were identified using the IDL[12] *RegionGrow* function, setting the in-function region identification limits from 0.04 to 1 and −0.04 to −1, which correspond to redshifted and blueshifted events, respectively. The 3D nature of the data set meant that event pixels with no neighbors would have a value of 1/27. These thresholds were chosen to select any pixel which had at least two neighboring pixels labeled as an event, and reject any pixels which only had a single neighbor marked as an event, assuming them to be noise. Event times were calculated by taking the difference in time between the latest and earliest times of the pixels of each event. The results of this procedure can be found in Section 3 of this paper.

---
[12] Interactive Data Language.

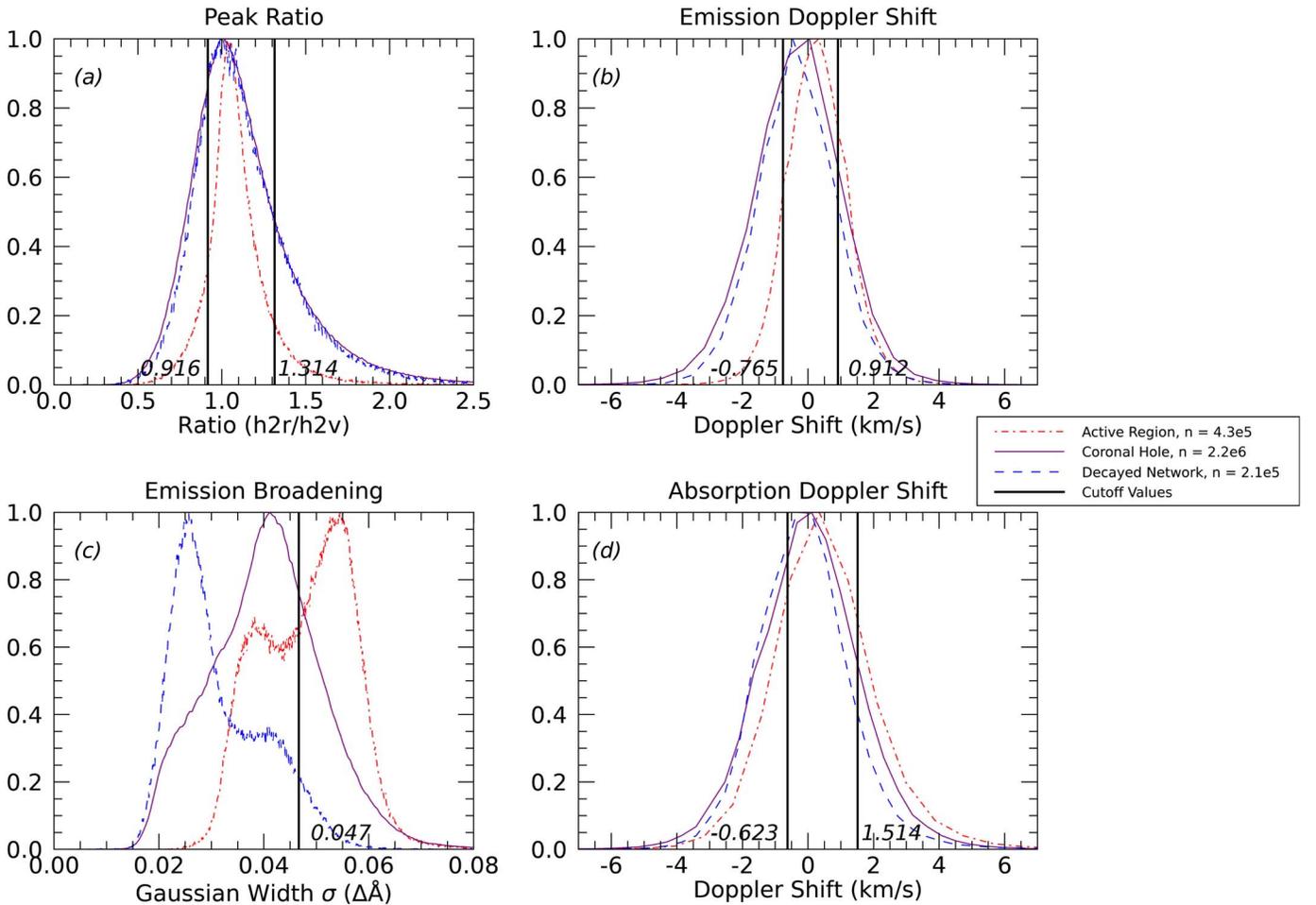

**Figure 3.** Distribution histograms for the spicule identification parameters: peak ratio (a), emission Gaussian Doppler shift (b), emission Gaussian broadening (c), and absorption Gaussian Doppler shift (d). Data corresponding to the active region, coronal hole, and decayed network are respectively represented by red dotted–dashed lines, purple solid lines, and blue dashed lines. The black vertical lines show the locations of the cutoff values for each parameter and the associated number provides the exact value.

### 2.6. The Magnetic Field Data

The interpolated HMI locations underneath the slit were ordered by absolute magnetic field strength and binned so that each bin contained 100 pixels (locations). For each bin, the number of pixels marked as an event over time at those 100 locations was calculated. Then the maximum possible number of events was calculated by multiplying the number of observations (time) by the number of locations (100). Those two numbers were compared to estimate what fraction of the time a given magnetic field strength would have a spicule. Using this method, each bin is weighted the same. Wider bins covering a larger range of field strengths correspond to lower resolution, while narrower bins covering a smaller range of field strengths correspond to higher-resolution information. This method of binning was chosen as a way to highlight the relative confidence or error in the results, and the number of values in each bin was chosen to make both low- and high-field-strength data visually easy to read.

### 3. Results and Discussion

#### 3.1. Analysing Individual Events

Our algorithm identified a total of 2021 spectra as events with a large variety of spectral shapes. This algorithm identifies spicule spectra which are not readily apparent to visual inspection, guided by the visually identifiable parameters set out by Rouppe van der Voort et al. (2015). Each positive event satisfies the requirements set out in Section 2.3. We refer to the Appendix for a sample of spectra identified as spicules.

##### 3.1.1. Potential Type I Spicules

While we identify individual Mg II spectra as spicules without additional spatial or temporal information, the temporal dimension can provide important information, as it was key to the original identification of spicules in Mg II in the form of RBEs and RREs (Rouppe van der Voort et al. 2015). When looking at the spectrum of a single location over time, some spectra display a sawtooth pattern at location of their central h3 absorption feature (see Figure 4(b) for an example). As time progresses, the h3 absorption feature starts strongly blueshifted, drifts to being strongly redshifted over the course of 4–10 minutes, and then jumps back to being blueshifted. This sawtooth pattern was first modeled in the Ca II H & K lines in bright grains by Rutten & Uitenbroek (1991) and then by Carlsson & Stein (1997), who showed that the phenomenon was caused by shocks propagating in the chromosphere. De Pontieu et al. (2004), Hansteen et al. (2006), and De Pontieu et al. (2007) later confirmed that this shock pattern was influenced by the Sun's 3–5 minute p-modes and was related to dynamic fibrils, which became known (retroactively after the discovery of Type II spicules by De Pontieu et al. 2007) as Type I spicules. Skogsrud et al. (2016) provides additional details on the spectral evolution of Type I spicules in Mg II k and other chromospheric and TR observations. Figure 4(b) displays a λ–time plot which displays a sawtooth pattern from the active region data set. For the locations in our data sets which display this magnetoaccoustic shock pattern, nearly all redshifted events appear minutes just before the central reversal jumps back to blue, and nearly all blueshifted events are located in the minutes just after the jump. Not all times before and after the Doppler jump are marked as events using this identification method. Many do not pass the broadening and/or emission Doppler shift requirement, and instead simply have the respective red/blue peak entirely suppressed. These spectra also deviate most from the assumed double-Gaussian line formation, for which the fitting method of Schmit et al. (2015) often produces fits with restrictively high reduced-$\chi^2$ values.

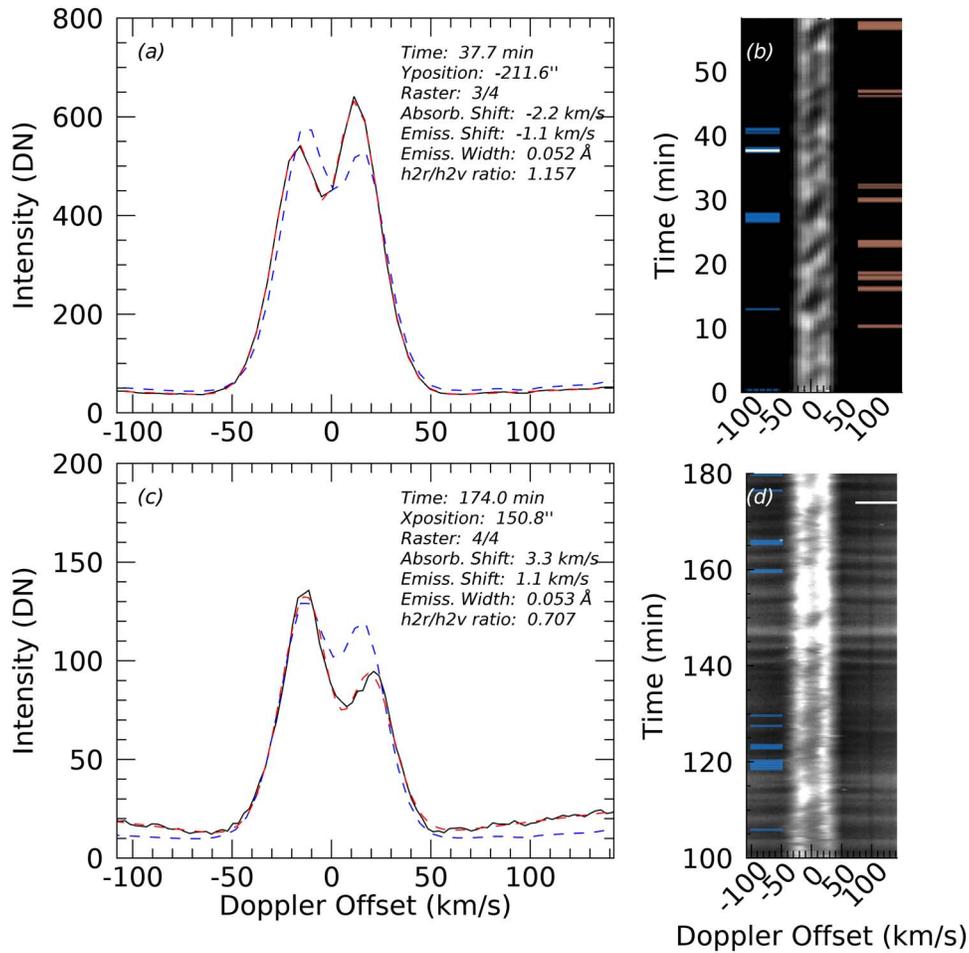

**Figure 4.** Comparison between a potential Type I and Type II spicule. Subplot (a) shows the spectrum of a potential Type I spicule from the active region data set (black), its modeled Gaussian fit (red dashed), and a median reference Mg II profile (blue dashed) with arbitrary intensity units. The spectrum's time and location are listed at the top right along with the Gaussian fit parameters used in this analysis. Subplot (b) shows the location's spectrum over time with the scale adjusted to highlight the central feature. The blue lines on the left indicate when the location is identified as a blueshifted spicule, and the red lines on the right indicate redshifted spicule times. The white line corresponds to the profile from (a). Subplots (c) and (d) are formatted the same as (a) and (b), but show a redshifted potential Type II spicule from the coronal hole data set and (d) is scaled to highlight the wings.

Currently, we do not have a method to identify automatically if a series of spectra has a sawtooth central inversion pattern, but we visually estimate that roughly 60% of the active region event spectra show such a pattern, while it is only about 33% for the coronal hole data set. Assuming that the sawtooth pattern is indicative of a Type I spicule, this agrees with De Pontieu et al. (2007) and Pereira et al. (2012) that Type I spicules are more likely to occur near active regions.

*3.1.2. Potential Type II Spicules*

Most λ–time plots in these data sets (especially in the decayed network or coronal hole data sets) do not display a strong or regular sawtooth pattern but instead simply show unstructured variation in both the central reversal and in the wings. These spectra often show small central excursions and wing brightenings similar to the RBEs and RREs described in Rouppe van der Voort et al. (2015), but are rarely as extreme or obvious as shown in that work. Figure 4(d) displays a λ–time plot which displays potential Type II spicules from the coronal hole data set. Agreeing with Pereira et al. (2012), we find that spectra with Type II–like behavior predominantly occur in the coronal hole data set.

Table 3
Event Location Statistics

| Data set | Number of Redshift Events | Number of Blueshift Events | Redshift Event Rate (arcsec$^{-2}$ min$^{-1}$) | Blueshift Event Rate (arcsec$^{-2}$ min$^{-1}$) | Redshift Fill Factor | Blueshift Fill Factor |
|---|---|---|---|---|---|---|
| Coronal Hole | **592** | **929** | 0.055 | 0.086 | 4.4% | 4.5% |
| Decayed Network | 58 | 70 | 0.023 | 0.028 | 0.79% | 0.59% |
| Active Region | 209 | 163 | 0.084 | 0.065 | 5.7% | 2.2% |

### 3.2. Event Detection Statistics

Table 3 shows the number of spicules, the percentages of pixels (in space and time) with events, and the rate of identified events over time for all three data sets. We believe this algorithm undercounts the number, size, and duration of events with a suppressed h2 peak, and overcounts events which may barely exceed the required thresholds.

When identifying spicules in an enhanced network region using SST Hα and Ca II K, Bose et al. (2021) identified 19,643 RBEs and 14,650 RREs over the course of 97 minutes in a data set that was roughly 54″ × 50″. This corresponds to a rate of 0.056 RREs minute$^{-1}$ arcsec$^{-2}$ and 0.075 RBEs minute$^{-1}$ arcsec$^{-2}$. We find similar redshifted event rates in our coronal hole data set, but slightly higher blueshifted rates. The active region data set shows similar event rates overall, but the distribution between redshifted and blueshifted events is weighted toward red. In comparison, we identify events in the decayed active network at a rate of 0.023 RREs minute$^{-1}$ arcsec$^{-2}$ and 0.028 RBEs minute$^{-1}$ arcsec$^{-2}$, which are about 50%–70% less than in the active region.

When analysing an extremely quiet-Sun region displaying Type II spicules, Henriques et al. (2016) found a filling factor of 1% and 1.3% for RREs and RBEs, respectively, using Hα, which are nearly double our findings for the decayed active network data.

While in general spicules occur ubiquitously on the Sun, they occur with high density in and near active regions. (De Pontieu et al. 2004, 2007; Bose et al. 2019, 2021). This work agrees with these findings, as can be seen in Figure 5(a) in the active region data set, which show that spicules are more likely to be found near regions of high magnetic field strength in general. Events in an active region were more likely to occur in the bright strongly magnetic areas in the SJI 1400 data while almost ceasing to exist in the darker internetwork regions, which have low magnetic fields. The coronal hole data set also displays a high density of spicules. Events tended to recur in the same location over time. In the active region data set, the maximum time a single location was marked as an event was 32% in brighter active regions while the minimum time was 0% in the darker internetwork areas. In the coronal hole data set this maximum recurrence density was 32% while in the decayed active network data set, it was 23%.

Of particular note, a small upflowing jet located near helioprojective Cartesian coordinates Solar-X and Solar-Y = [−190″, −265″] in the active region data set directed upflowing material across the slit location for the entire duration of the observation. The upflowing material was identified as a series of long-lasting blueshifted events at −270″ in Figures 5(a), (b). The jet is an example of a type of event which can be identified by this method but which is obviously not a spicule. Multiple detections of this jet lasting longer than 300 s were identified and noted as anomalous. These features can be filtered out by using stronger peak ratio requirements and setting a maximum temporal constraint, but were kept in this analysis as an example.

Another anomalous feature originally detected in the coronal hole data set was a high density of events between Solar X coordinates (43″, −53″) at the location of the masking box in Figure 5(d). Upon examining the HMI data sets between the beginning and the end of the IRIS observation period, we found a difference of more than 200 G at this location. This is a region with some small-scale negative magnetic fields within the larger positive coronal hole. Events detected here are not due to spicules, but instead may be siphon flows connected by small magnetic loops within the overall coronal hole structure. We have manually removed all events identified in this region from our analysis. Partially because of this feature, we believe the algorithm may overcount small events.

### 3.3. Event Time Statistics

The IRIS rasters repeated at 12.7 s, 9.4 s, and 21.6 s seconds for the active region data set, decayed active network data set, and coronal data set, respectively. These values define the lower bounds for event duration. To qualify as an event and to remove spurious detections, events were required to have some extent in either time or space. Pixels which had no neighbors in either time or space were removed as noise, which eliminated most of the shortest-duration identifications.

Event duration for our data sets can be found in Table 4. These durations are longer than the median duration of 27.2 s identified by Bose et al. (2021). Bose et al. (2021) also found that 98% of events had lifetimes shorter than 200 s. In comparison, we find that only 94% of our active region events have durations shorter than 200 s and 88% of the coronal hole events had durations shorter than 200 s. This was unexpected because Type II spicules have shorter durations and are more often found in coronal holes than active regions. The longer durations are likely influenced by the high density of identified events, causing smaller events nearby in time to be counted as a single longer event. This may also be influenced by the increased raster duration of the coronal hole data set (21.6 s compared to 12.7 s and 9.4 s), which is known to affect mean observed event durations (Pereira et al. 2013).

Other works (Rouppe van der Voort et al. 2009; Pereira et al. 2012; Sekse et al. 2012; Kuridze et al. 2016) found longer durations for RBEs and RREs ranging from 24 to 400 s. Bose et al. (2021) suggests that this may be because some of these works focused on single wavelength positions far in the blue or red wing of the Hα or Ca II line profiles.

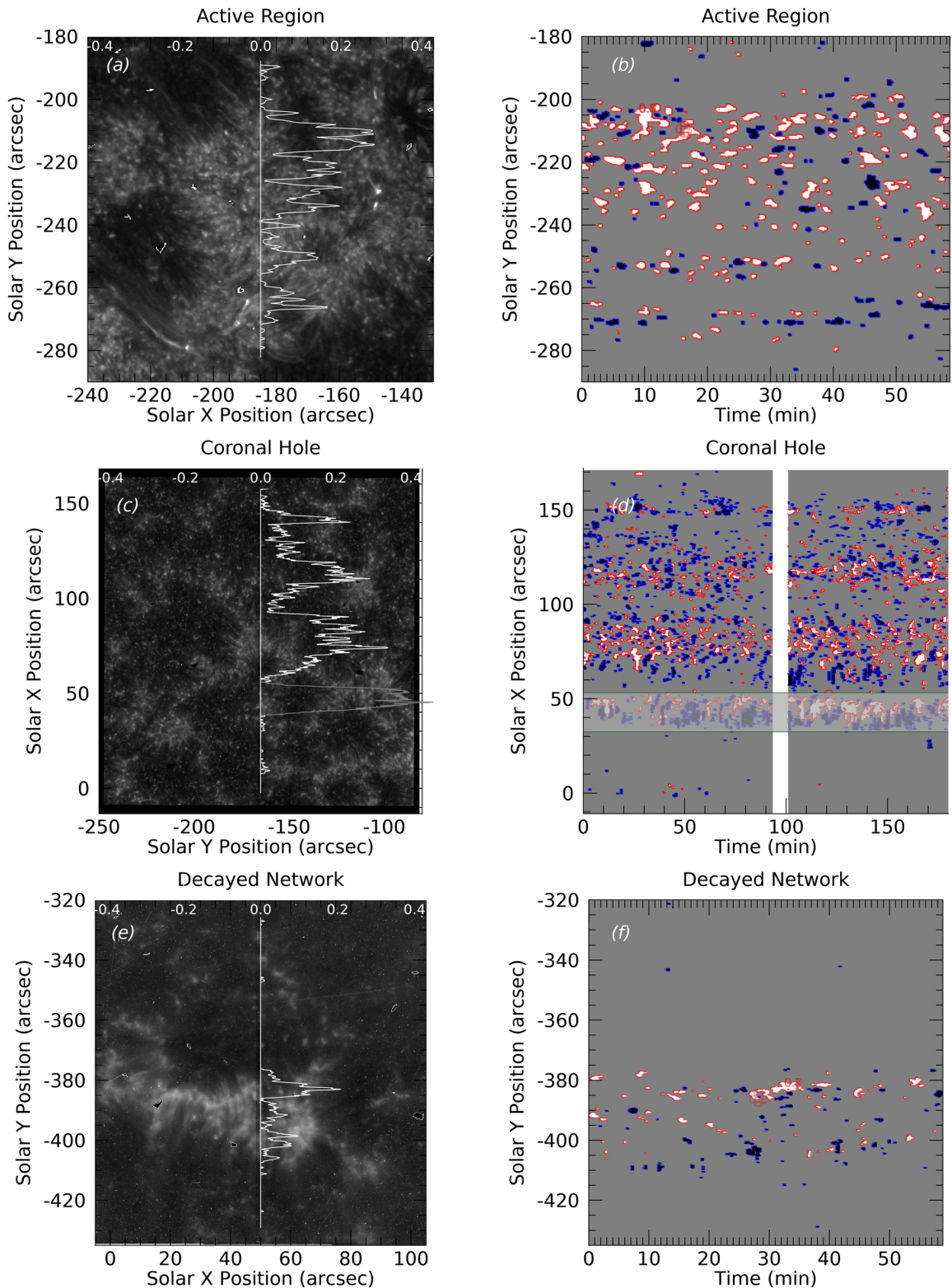

**Figure 5.** Event location statistics for the active region (panels (a) and (b)), coronal hole (panels (c) and (d)), and decayed active network (panels (e) and (f)) data sets. Left column: event recurrence along the slit plotted against an IRIS SJI 1400 image for context. The vertical white lines show the slit location while the variable lines show what fraction of time a given location is identified as an event. Right column: maps of event identification over time. The white/red regions are redshifted while the black/blue regions are blueshifted. Each plot shows events from the first raster position over time. Data contaminated by the South Atlantic Anomaly have been removed (vertical white strip in (d)). Events underneath the box covering Solar $X$ 43″–53″ have also been removed from this paper's statistics due to suspect magnetic activity.

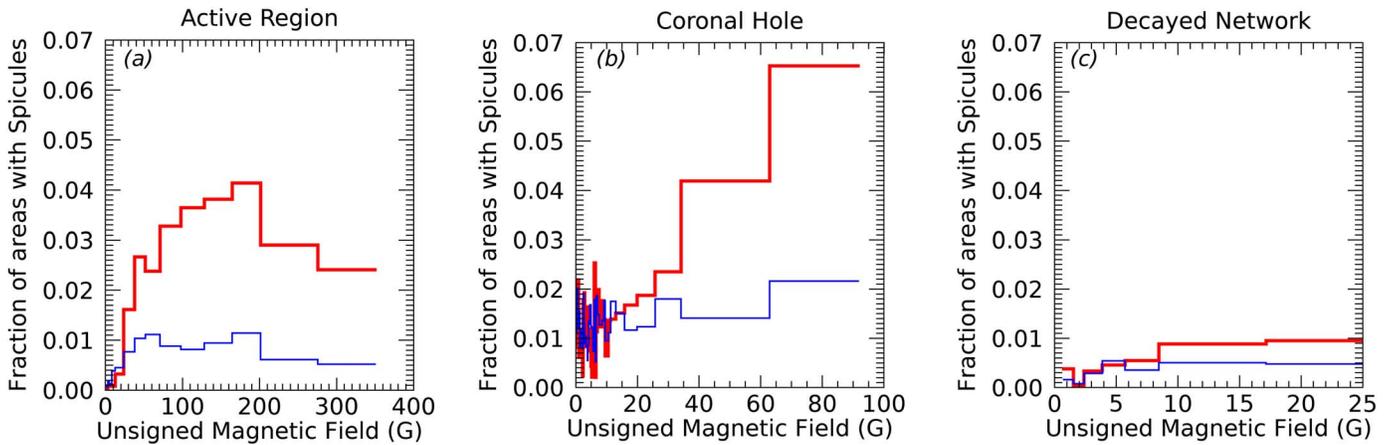

**Figure 6.** Event occurrences with relation to the absolute magnetic field strength for the active region (a), coronal hole (b), and decayed active region (c) data sets. The thicker red line corresponds to redshifted events while the thinner blue line corresponds to blueshifted events. Wider bins indicate lower *B*-field resolution and less homogeneity. The final bin in subplot (c) extends to 50 G, but has been truncated for plot readability.

**Table 4**
Identified Event Time Statistics

| Data set | Median Event Duration (s) | Mean Duration and Standard Deviation of Red Events (s) | Mean Duration and Standard Deviation of Blue Events (s) | Median Duration of Red Events (s) | Median Duration of Blue Events (s) |
|---|---|---|---|---|---|
| Coronal Hole | 76 | $122 \pm 118$ | $101 \pm 81$ | 76 | 76 |
| Decayed Network | 38 | $64 \pm 57$ | $46 \pm 28$ | 39 | 38 |
| Active Region | 57 | $90 \pm 86$ | $65 \pm 50$ | 57 | 44 |

### 3.4. Events in Relation to Magnetic Field Strength

When examining distribution of events with respect to the magnetic field strength in Figure 6, we find that locations with an absolute field strength below 20 G have only a 0.5%–2% chance of being identified as an event at any given time across all data sets, while locations with an absolute field strength greater than 20 G have a 2%–6% chance of being identified as an event at any given time. From the data sets examined, it appears that spicule activity in active regions peaks between 100 and 200 G at 4%–6%, but a lack of field strengths above 200 G precludes analysis at stronger magnetic fields. When reading Figure 6, it should be noted that each bin contains exactly 100 HMI pixel locations, so wider bins represent less homogeneous data and thus lower *B*-field resolution. From this, we suggest that without foreknowledge of spicular activity, a mission looking in Mg II or Ly$\alpha$ should observe active regions with field strengths between 100 and 200 G.

We wish to caution readers that there are several factors which are not considered in this magnetic field comparison. The projection effects due to the height difference between where the Mg II h core forms and where the underlying HMI measurements are made are similar to or just slightly greater than the HMI spatial resolution, which may result in a slight misalignment of data sets. Additionally, the spatial extent of spicules means that some footpoints may be up to several arcseconds from their observed Mg II h locations. Finally, the HMI resolution in this analysis is not sufficient to characterize fully all magnetic fields, which may modulate spicule activity. These combined mean that the Mg II h spicule locations are likely offset from the HMI magnetic field data by some small unknown and irregular amount. Because of this, the magnetic field analysis should not be used for anything more specific than general trends.

### 4. Model Limitations

As with the initial design of any such algorithm, there will be false positives and false negatives in this method. Bose et al. (2019) provides a thorough analysis of Mg II spicule spectral line formation both in modeling and comparative observations with H$\alpha$ and Ca II, which strongly supports our identification method. Many of these false positives and negatives can be eliminated with additional tuning using a comparison to co-observed data sets.

One of the primary assumptions in this method is that the Mg II line can be fitted by the combination of a positive and a negative Gaussian. While this assumption holds true for most data, spectra with highly Doppler-shifted central inversions break this assumption. Because of this, fits for the most extreme spectra either have high reduced-$\chi^2$ values or cannot be fit at all. When discarding poorly fitted spectra, the reduced-$\chi^2$ cutoff is particularly important because the most extreme spicule spectra are most likely to have a poor fit. In the active region data set, a reduced $\chi^2 < 5$ discards 30% of the fits, a reduced $\chi^2 < 10$ discards 18% of fits, and a reduced $\chi^2 < 20$ discards 13% of fits. This analysis used a reduced $\chi^2$ of 10. Because this model struggles to fit the more extreme Doppler-shifted spectra and more extreme spectra occur more often in active regions, we can assume that we are undercounting the number of extreme spicules, particularly in and around active regions.

When examining the $\lambda$–time plots of individual locations that feature the sawtooth pattern from Section 3.1 we note that while events are most often identified at the extremes of the sawtooth pattern (which is indicative of Type I spicules), not all extremes are identified as events. While we do not assume that all extremes must be events, we did review the spectra at the extremes of the sawtooth pattern to understand the cause of these potential false negatives. Most of these spectra observed either had a high reduced $\chi^2$, or failed to display both a broadening and a Doppler shift of the Gaussian emission feature. The lack of broadening and Doppler shift of the Gaussian emission feature may also be due to the extreme absorption of the Doppler-shifted plasma above, in the same way that one of the k2 peaks may be entirely suppressed.

As stated in Section 2.3, each threshold value was chosen such that it identified half of the spectra for each parameter as a potential event, and combining all requirements further reduced the spectra selection. This method for determining parameters must be tuned by comparing events identified in Mg II to co-observed events in H$\alpha$ or Ca II as laid out in Section 5. Proper tuning may be able to remove anomalous features such as the jet and potential siphon flows mentioned in Section 3.2.

Figure 3(c) shows that the calculated emission broadening for the active region and decayed network data sets do not actually have a Gaussian distribution, but are instead bimodal. The width of the emission feature is correlated with the underlying magnetic field strength, so this selection parameter and our results will be biased toward strong magnetic field strengths. This is expected, as the width of the emission Gaussian is a measure of the broadening overall, which is automatically correlated with the total intensity. We have chosen to use a single statistical modifier across all results regardless of the Solar magnetic regime of the data set, as doing otherwise would require observations to be definitively categorized into a magnetic regime before analysis. This bias will be addressed in a future paper described in Section 5.

## 5. Future Investigations

Now that a large sample of spicule spectra has been identified, we can use $K$-means clustering similar to Bose et al. (2019, 2021) to describe Mg II spicule profiles better beyond this double-Gaussian approach. $K$-means clustering can be used to identify the most common spectral profiles in the data set and examine the selection bias that high reduced-$\chi^2$ profiles induce in this method. Additionally, this method could be further refined by comparing co-observed H$\alpha$ and Mg II data sets for a direct comparison between event locations in H$\alpha$ or Ca II and Mg II. Since most spicule activity has been identified using H$\alpha$ or Ca II so far, this comparison could provide a validation for Mg II event locations as well as an opportunity to tune the cutoff values in this analysis. Finally, co-observations of H$\alpha$ and Ly$\alpha$ could be used to understand the statistical evolution of a spicule better as it heats to chromospheric or (likely) coronal temperatures.

## 6. Conclusion

In this analysis we suggest a method for identifying spicule spectra exclusively using Mg II spectral profiles. We make the simplifying assumption that an Mg II h spectrum can be modeled using a positive emission Gaussian plus a negative absorption Gaussian. From there, we use the Doppler shift and width of the emission Gaussian along with the Doppler shift of the absorption Gaussian and ratio between the h2 peaks to identify 2021 spicule events. We perform this analysis for three data sets spanning a range of solar magnetic environments: active region, decayed active region, and coronal hole, using both sit-and-stare observations and high-cadence raster observations. The reduced temporal coverage and additional spatial coverage of the rasters do not meaningfully change the results of this paper.

Both Type I and Type II spicules are identified during this analysis, with Type I events occurring more often in the active region data set. This analysis identifies redshifted and blueshifted events with a fill factor of 5.7% and 2.2%, respectively, for the active region data set, 4.4% and 4.5% for the coronal hole data set, and decreasing to 0.79% and 0.59% for the decayed active network data set. This decrease in event identification is likely related to the lack of absolute magnetic field strengths above 20 G. The decayed active region fill factor is about half the factor that Henriques et al. (2016) found in a quiet-Sun region using H$\alpha$ observations.

When analysing the normalized event rate over time, we identify redshifted and blueshifted events in an active region at a rate of 0.084 arcsec$^{-2}$ minute$^{-1}$ and 0.065 arcsec$^{-2}$ minute$^{-1}$, respectively, which are likely undercounts. This is comparable to the overall rates found by Bose et al. (2021), though that study found more blueshifted events than redshifted events. The coronal hole data set yields events at a similar rate (though skewed toward blueshifted), which is likely an overcount, while the decayed active network yields a much lower rate.

When compared to magnetic field strengths, we find that 0%–2% of locations with an absolute magnetic field strength less than 20 G had identified events, while locations with absolute magnetic field strengths of 20–200 G were identified as events 2%–6% of the time, with active region event identifications peaking around 4% between 100 and 200 G. Coronal hole events appeared to increase steadily above 15 G but the statistics become unreliable above 40 G. The data sets analysed displayed few magnetic field strengths above 200 G, which preclude any analysis at higher strengths. From this we conclude that without foreknowledge of spicular activity, a mission looking in Mg II or Ly$\alpha$ should observe active regions with field strengths between 100 and 200 G.

This analysis struggles to identify the most extreme spicule spectra for which one emission peak is entirely suppressed. A more thorough $K$-means clustering method combined with H$\alpha$ co-observations is proposed to supplement this method.

In addition to presenting a new method for identifying spicules in Mg II without using time-dependent methods, this analysis also presents 2021 spicule spectra identified algorithmically, 24 of which are shown in the Appendix. Many previous examples of Mg II spicule spectra provided only the most extreme spectral examples, and these new spectra may be used to examine a wider variety of Mg II spicule spectral shapes.

V.L.H., P.C.C., D.S., A.D., and V.P. would like to acknowledge support from NASA grant 80NSSC22K0576 (SNIFS; PI: Chamberlin) and the NASA H-FORT program.

IRIS is a NASA small explorer mission developed and operated by LMSAL with mission operations executed at NASA Ames Research Center and major contributions to downlink communications funded by ESA and the Norwegian Space Centre.

The SDO/HMI data are courtesy of NASA/SDO and the HMI science teams.

S.B. and V.P. gratefully acknowledges support from NASA contract NNG09FA40C (IRIS).

R.O.M. would like to thank the Science and Technologies Facilities Council (UK) for the award of an Ernest Rutherford Fellowship (ST/N004981/2).

# Appendix
## Set of Identified Spicule Spectra

Figures 7, 8, and 9 provide a random selection of spicule spectra identified through this analysis. Data from Figure 7 come from the active region data set, while data from Figures 8 and 9 come from the decayed active network data set and the coronal hole data set respectively. The purpose of these figures is to provide a variety of spectral examples to highlight the diverse appearances of a spicule spectrum. In addition to spectral shape, the associated lambda-time plot is included which may provide information on whether the example displays a Type I or Type II spicule. The exact location and time of the spicule are provided at the top right of each spectrum, along with its derived spectral parameters.

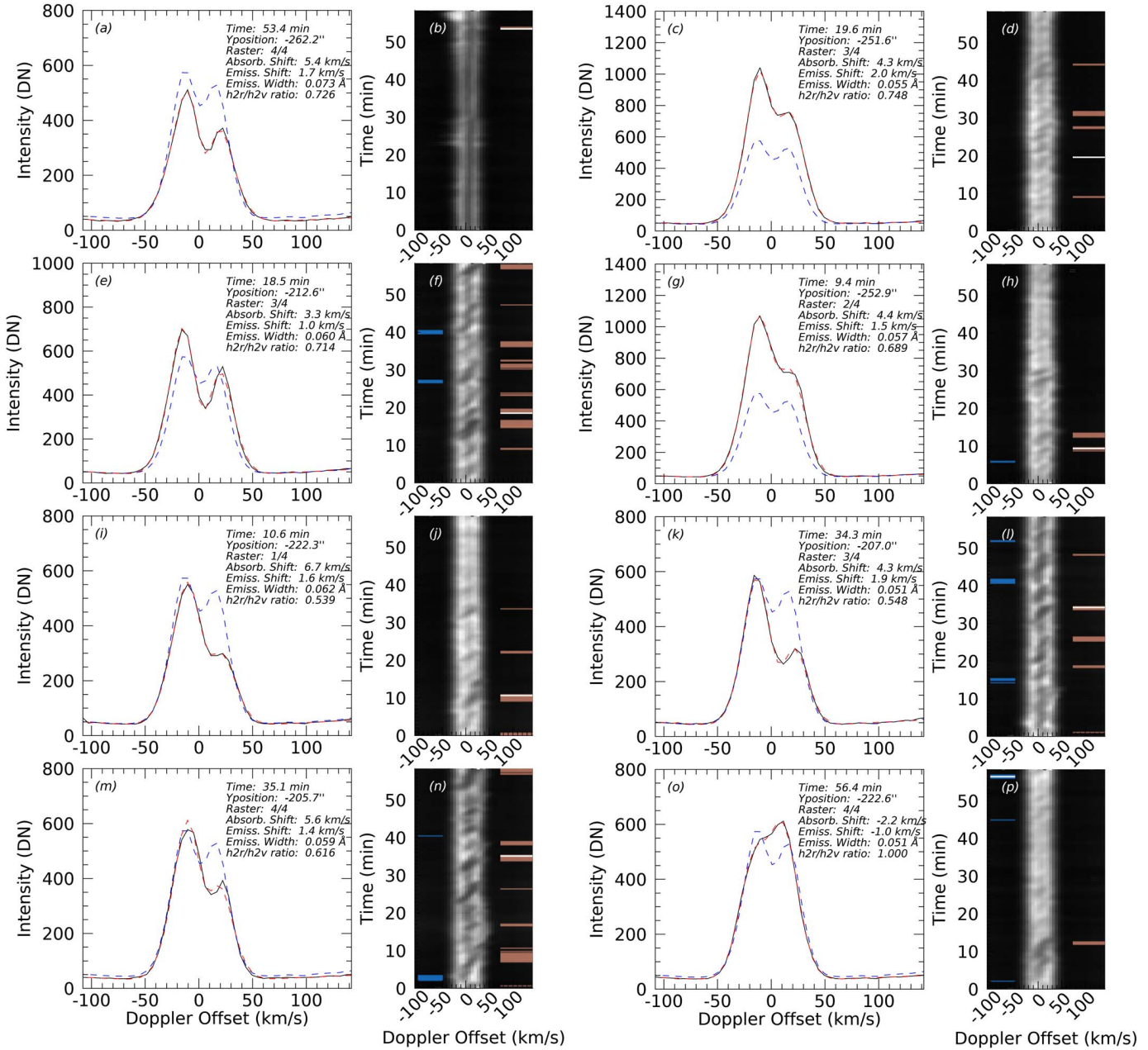

**Figure 7.** Random spectrum samples of identified spicules from the active region data set. Subplot (a) shows a spectrum of potential Type I spicules from the active region data set (black), its modeled Gaussian fit (red dashed), and a typical Mg II profile (blue dashed) with arbitrary intensity units. The spectrum's time and location are listed at the top right, as well as the Gaussian fit parameters used in this analysis. Subplot (b) shows that same location's spectrum over time with the scale adjusted to highlight the central feature. The blue lines on the left indicate when the location is identified as a blueshifted spicule, and red lines on the right indicate redshifted spicule times. The white line corresponds to the profile from (a). The remaining subplot pairs are formatted the same as (a) and (b).

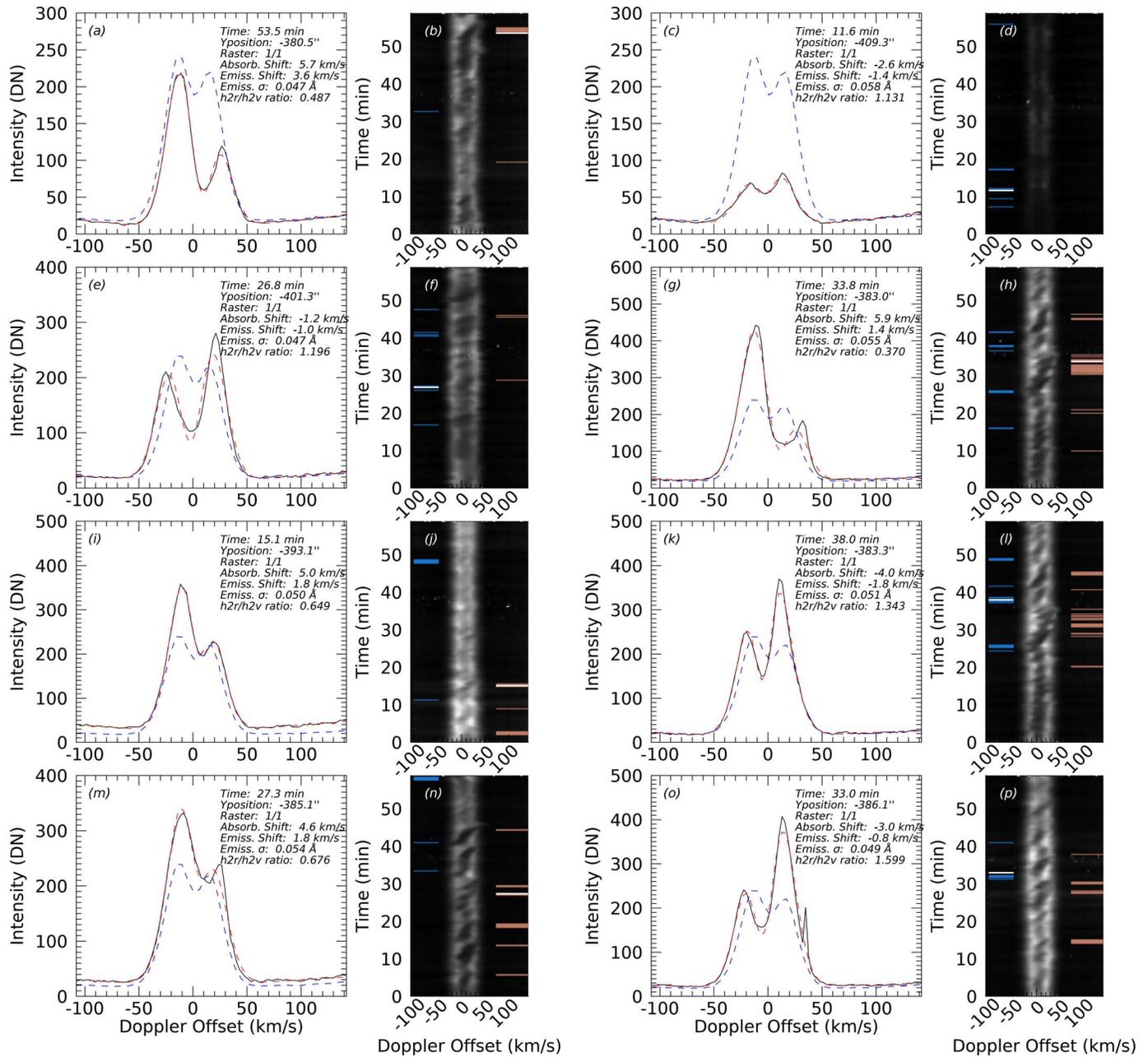

**Figure 8.** Random spectrum samples of identified spicules from the decayed active network data set. The figure features and formatting are the same as in Figure 7.

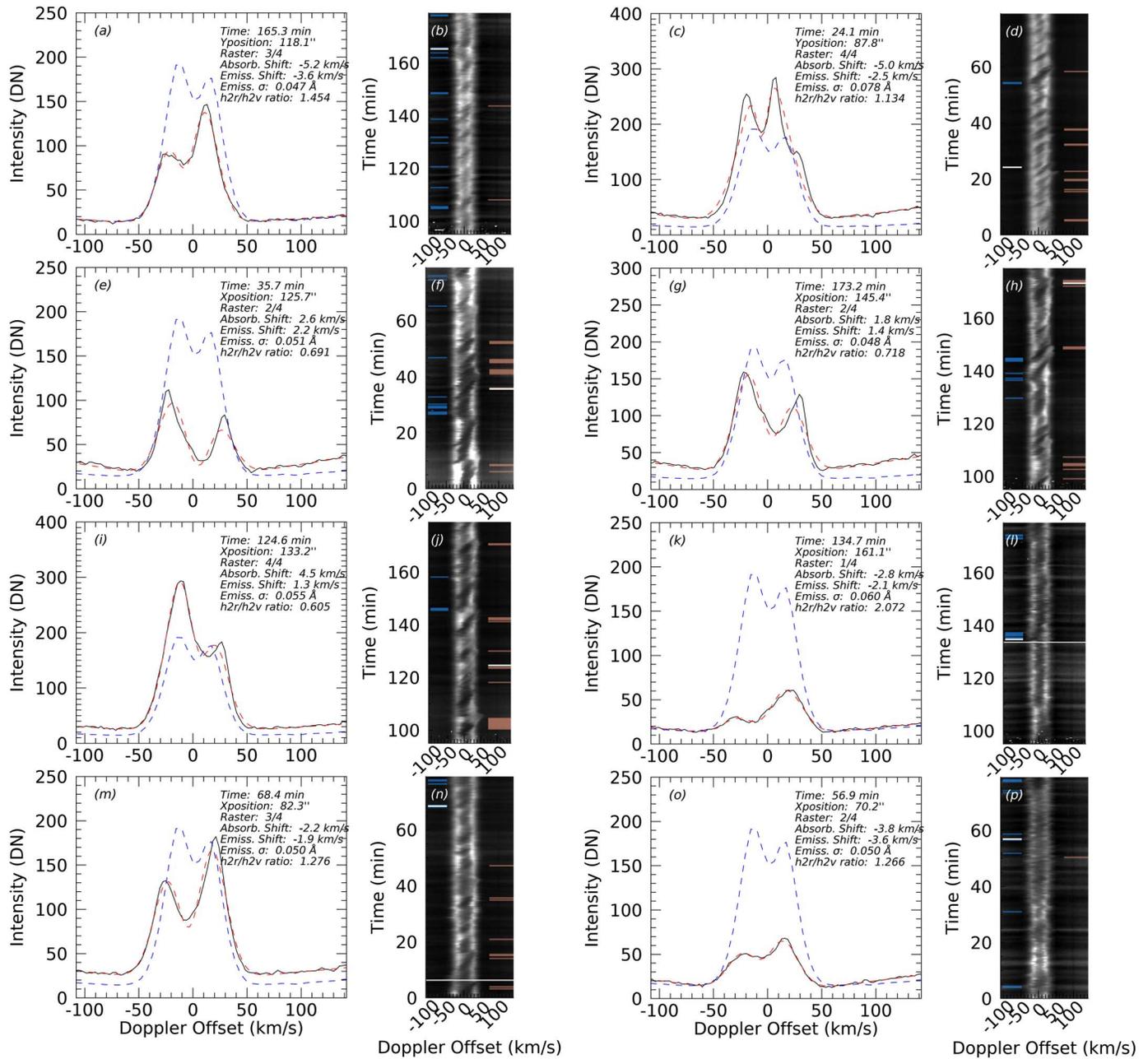

**Figure 9.** Random spectrum samples of identified spicules from the coronal hole data set. The figure features and formatting are the same as in Figure 7.

## ORCID iDs

Vicki L. Herde https://orcid.org/0000-0001-9139-8939
Phillip C. Chamberlin https://orcid.org/0000-0003-4372-7405
Don Schmit https://orcid.org/0000-0002-9654-0815
Souvik Bose https://orcid.org/0000-0002-2180-1013
Adrian Daw https://orcid.org/0000-0002-9288-6210
Ryan O. Milligan https://orcid.org/0000-0001-5031-1892
Vanessa Polito https://orcid.org/0000-0002-4980-7126